\documentclass[aps,pra,reprint,amsmath,amssymb,superscriptaddress,nobibnotes]{revtex4-1}
\usepackage{ulem,bm}
\usepackage{amsmath, upgreek, amssymb, graphicx, paralist, gensymb,epstopdf}
\usepackage[colorlinks, linkcolor=blue, citecolor=blue, urlcolor=blue]{hyperref}
\usepackage{array,epsfig,amsmath,amssymb}
\usepackage{graphicx}
\usepackage{color}

\def\bra#1{\langle #1|}
\def\ket#1{|#1 \rangle}

\begin{document}

\title{Global triplewise information trade-off in quantum measurement}

\author{Seongjin Hong}
\affiliation{Center for Quantum Information, Korea Institute of Science and Technology (KIST), Seoul, 02792, Korea}

\author{Yong-Su Kim}
\affiliation{Center for Quantum Information, Korea Institute of Science and Technology (KIST), Seoul, 02792, Korea}
\affiliation{Division of Nano and Information Technology, KIST School, Korea University of Science and Technology, Seoul 02792, Korea}

\author{Young-Wook Cho}
\affiliation{Center for Quantum Information, Korea Institute of Science and Technology (KIST), Seoul, 02792, Korea}
\affiliation{Department of Physics, Yonsei University, Seoul 03722, Korea}

\author{Jaewan Kim}
\affiliation{School of Computational Sciences, Korea Institute for Advanced Study, Seoul 02455, Korea}

\author{Seung-Woo Lee}
\email{swleego@gmail.com}
\affiliation{Center for Quantum Information, Korea Institute of Science and Technology (KIST), Seoul, 02792, Korea}

\author{Hyang-Tag Lim}
\email{hyangtag.lim@kist.re.kr}
\affiliation{Center for Quantum Information, Korea Institute of Science and Technology (KIST), Seoul, 02792, Korea}
\affiliation{Division of Nano and Information Technology, KIST School, Korea University of Science and Technology, Seoul 02792, Korea}

\date{\today} 

\begin{abstract}
State disturbance by a quantum measurement is at the core of foundational quantum physics and constitutes a fundamental basis of secure quantum information processing. While quantifying an information-disturbance relation has been a long-standing problem, recently verified reversibility of a quantum measurement requires to refine such a conventional information trade-off toward a complete picture of information conservation in quantum measurement.
Here we experimentally demonstrate complete trade-off relations among all information contents, i.e.,~information gain, disturbance and reversibility in quantum measurement.
By exploring various quantum measurements applied on a photonic qutrit, we observe that the information of a quantum state is split into three distinct parts accounting for the extracted, disturbed, and reversible information. We verify that such different parts of information are in trade-off relations not only pairwise but also globally all-at-once, and find that the global trade-off relation is tighter than any of the pairwise relations. 
Finally, we realize optimal quantum measurements that inherently preserve quantum information without loss of information, which offer wider applications in measurement-based quantum information processing. 
\end{abstract}

\maketitle

Quantum measurement is at the heart of foundational quantum physics \cite{Heisenberg} and plays a major role to 
readout information from a quantum state in quantum technologies \cite{Wiseman09,Kurt,QIQM,teleportation,MBQC}.
However, since a quantum measurement inevitably disturbs the measured quantum system, the amount of extracted information from a quantum state has a certain limit against the state disturbance \cite{Groenewold71,Lindblad72,Ozawa86,Fuchs96}. 
A long-standing wisdom of this has been, `the more information of a quantum state is extracted by a quantum measurement the more the state is disturbed.' Such a trade-off relation has both fundamental and practical importance in establishing a basis of secure quantum information processing \cite{Gisin2002,BB84}. Numerous studies have been made to quantitatively verify a trade-off between information gain and state disturbance in quantum measurement \cite{Fuchs01,Banaszek01,Banaszek02,Dariano03,Sacchi06,Buscemi08,Luo10,Berta14}.

However, series of recent works observed that a quantum state disturbed by a quantum measurement weakly interacting with the measured system can be faithfully recovered by a post-measurement operation 
\cite{Ueda92,Chen14,Royer95,Ueda96,Jordan10,Terashima11,Terashima16,Cheong12,Koashi99,Terashima03,Korotkov06,YSKim09,Korotkov10,YSKim11,Katz08,Schindler13,JCLee11,HTLim14b}. 
This implicates that a part of total information remains and allows us to recover the original quantum state after the measurement. Accounting for such a reversible information, the reversibility of quantum measurement has been quantified and rigorously analyzed as an additional information content in quantum measurement \cite{Jordan10,Terashima11,Terashima16,Cheong12,HTLim14}.
Reversing quantum measurement has been realized in various physical platforms \cite{Katz08,Chen14,Schindler13,YSKim09,YSKim11,JCLee11,HTLim14b,HTLim14}, and applied for quantum error correction \cite{Koashi99}, gate operation \cite{Terashima03,Korotkov06}, and decoherence suppression \cite{YSKim09,Korotkov10,YSKim11,HTLim14b}.
Verifying trade-off relations encompassing all information contents, i.e., information gain, disturbance and reversibility in quantum measurement has thus become crucial particularly in applications to secure quantum technologies  \cite{Gisin2002,BB84}. 

This may be also important to establish a consistent picture of information conservation in quantum measurement. Penrose pointed out that many fundamental features in quantum mechanics remain veiled due to the inconsistency of the linearity and the measurement in quantum mechanics \cite{Penrose}. For example, while in the linearity of quantum mechanics (e.g., unitary evolution of quantum states) it is guaranteed that total information is preserved by the second law of thermodynamics, whether or when information is preserved in quantum measurement has remained unanswered to date.

In this work, we demonstrate complete trade-off relations in quantum measurement and realize quantum measurements preserving quantum information. We explore three information contents accounting for the extracted, disturbed and reversible information by performing different types of quantum measurements. To this end, we propose an experimental scheme for varying the type and strength of quantum measurement and apply this to path-encoded photonic qutrits.
We show that three information contents are quantitatively linked by a global triplewise trade-off relation as well as by three pairwise relations between each two of them, obeying the fundamental upper bounds derived in Refs.~\cite{Banaszek01,Cheong12,SWLEE20}. To our knowledge, this is the first experimental demonstration of the trade-off relation including information gain, disturbance, and reversibility of quantum measurement globally all-at-once. It shows that the global trade-off relation is tighter than any of the pairwise trade-off relations and reveals the emergence of the reversibility in multi-dimensional quantum measurements. Finally, we establish optimal quantum measurements inherently preserving quantum information, in the sense that all information is changed into another form by measurement without any missing part, which would find wider applications to measurement-based quantum information processing. 


\begin{figure}
\includegraphics[scale=0.22]{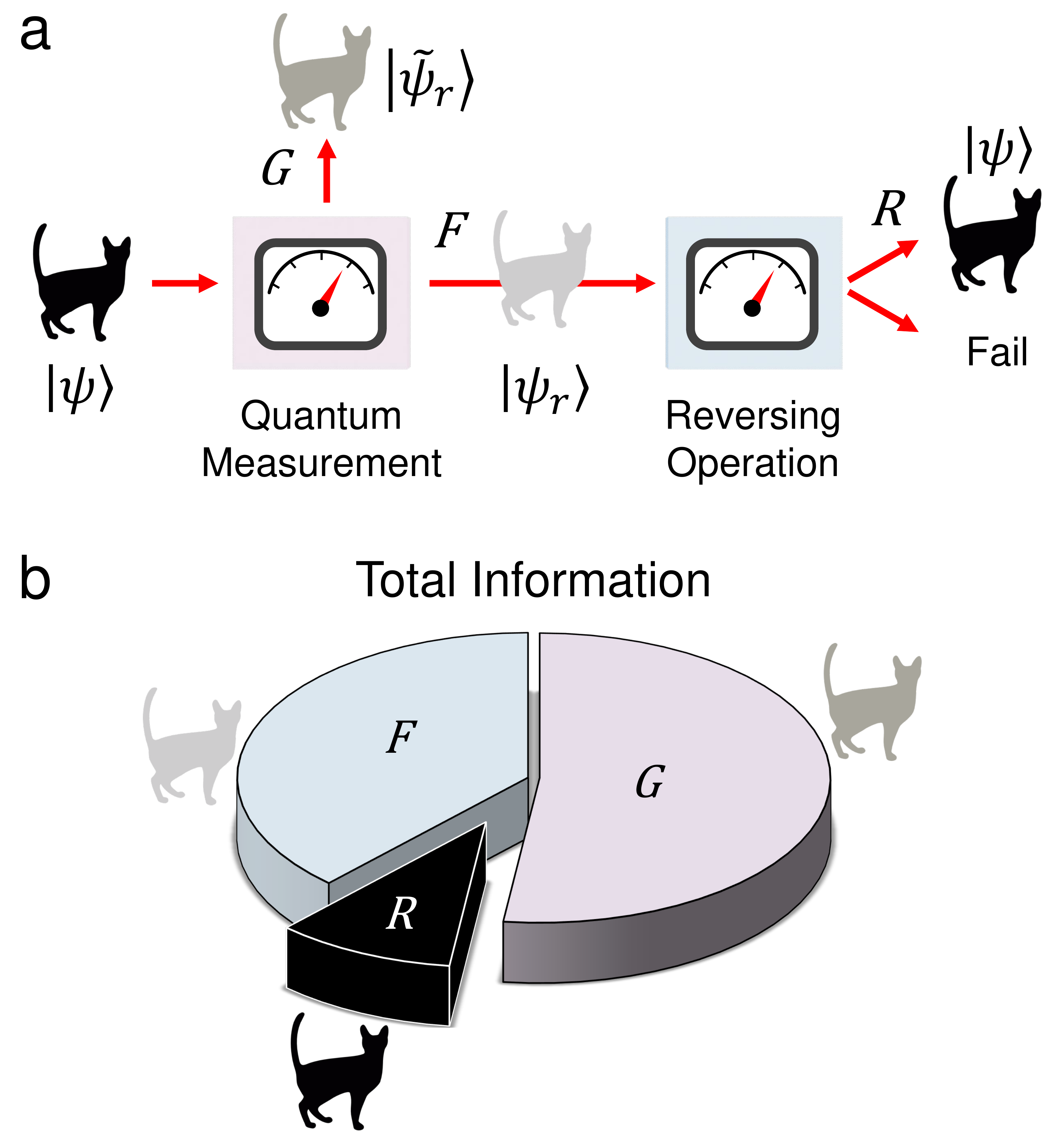} 
\caption{{\bf Three information contents in quantum measurement.} {\bf a,} Quantum measurement is performed to obtain information from an unknown input state $|\psi\rangle$. Information can be extracted by estimating the state based on the measurement outcomes (i.e.,~$|\tilde{\psi}_r\rangle$ for each outcome $r$), the amount of which is quantified as the information gain $G$. The input state is altered to $|\psi_r\rangle$ and the transmitted (undisturbed) information can be evaluated as the operation fidelity $F$. Then, a reversing operation can recover $\left | \psi \right\rangle$ probabilistically with the reversibility $R$. {\bf b,} The information of a quantum state is split into three parts, $G$, $F$, and $R$ by a quantum measurement. We demonstrate that $R$ fills the gap between $G$ and $F$ for the global tight trade-off relation.} 
\label{fig:fig1}
\end{figure}

Assume that an observer performs a quantum measurement to obtain information from a quantum state $\ket{\psi}$ (see Fig.~\ref{fig:fig1}{\bf a}). A quantum measurement can be described by a set of operators $\hat{M}_r$ satisfying the completeness condition $\sum_r\hat{M}_r^{\dag}\hat{M}_r=\hat{I}$. For each outcome $r$, the input state is altered to $\ket{\psi_r}=\hat{M}_r\ket{\psi}/{\sqrt{p(r,\psi)}}$ where $p(r,\psi)=\bra{\psi}\hat{M}_r^{\dagger}\hat{M}_r\ket{\psi}$, and the observer estimates the input state as $\ket{\widetilde{\psi}_r}$.
The amount of information extracted by the observer can then be quantified by the closeness between the initial and the estimated states $G= \int d \psi \sum_{r} p(r,\psi) |\langle \widetilde{\psi}_r | \psi \rangle|^2$ \cite{Banaszek01}, refereed to as the {\it information gain}. On the other hand, the {\em operation fidelity} defined by the closeness between the initial and the post-measurement states $F= \int d \psi \sum_{r} p(r,\psi) |\langle \psi_r | \psi \rangle|^2$ accounts for the amount of transmitted or undisturbed information. This in turn represents the amount of disturbance by $1-F$.

Assume that the observer then applies a reversing operation for each outcome $r$ attempting to restore the initial state $\ket{\psi}$ as illustrated in Fig.~\ref{fig:fig1}{\bf a}. The {\it reversibility} $R$ can  then be defined as the maximum success probability of the faithful recover of $\ket{\psi}$ after the measurement. The reversing operation is described by a set of operators $\{\hat{R}^{r}_l\}$ satisfying $\sum_l \hat{R}^{r \dag}_l\hat{R}^{r}_l =\hat{I}$. 
To reverse a quantum measurement $\hat{M}_r$, $\hat{R}^{r}$ should satisfy 
$\hat{R}^{r}\hat{M}_r\ket{\psi}\propto\ket{\psi},~\forall \ket{\psi}$.
Since a measurement operator can be written by $\hat{M}_r=\hat{V}_r\hat{D}_r\hat{U}_r$ in singular value decomposition with unitary operators $\hat{V}_r$ and $\hat{U}_r$ and a diagonal matrix $\hat{D}_r=\sum_{i=0}^{d-1}\lambda^r_i\ket{i} \bra{i}$ where $\lambda^r_i$ are the singular values in $d$-dimensional Hilbert space, the optimal reversing operator is given by $\hat{R}^{r}=\lambda^r_{\rm min}\hat{U}^{\dag}_r\hat{D}^{-1}_r\hat{V}^{\dag}_r$ with the smallest singular value $\lambda^r_{\rm min}$ among $\{\lambda^r_0,\ldots,\lambda^r_{d-1}\}$. 
The reversibility is then obtained as $R=\sum_{r}(\lambda^r_{\rm min})^2$ \cite{Cheong12}. In this representation, the information gain and operation fidelity can be also further evaluated as $G=(d+\sum_{r}(\lambda_{\rm max}^r)^2)/d(d+1)$ with the largest singular value $\lambda^r_{\rm max}$ and $F=(d+\sum_{r} (\sum_{i=0}^{d-1}\lambda_i^r)^2)/d(d+1)$, respectively \cite{Banaszek01}. Note that information gain is scaled in the range $1/d \leq G \leq 2/(d+1)$ and operation fidelity is scaled as $2/(d+1) \leq F \leq 1$. 

\begin{figure*}[t]
\includegraphics[scale=0.18]{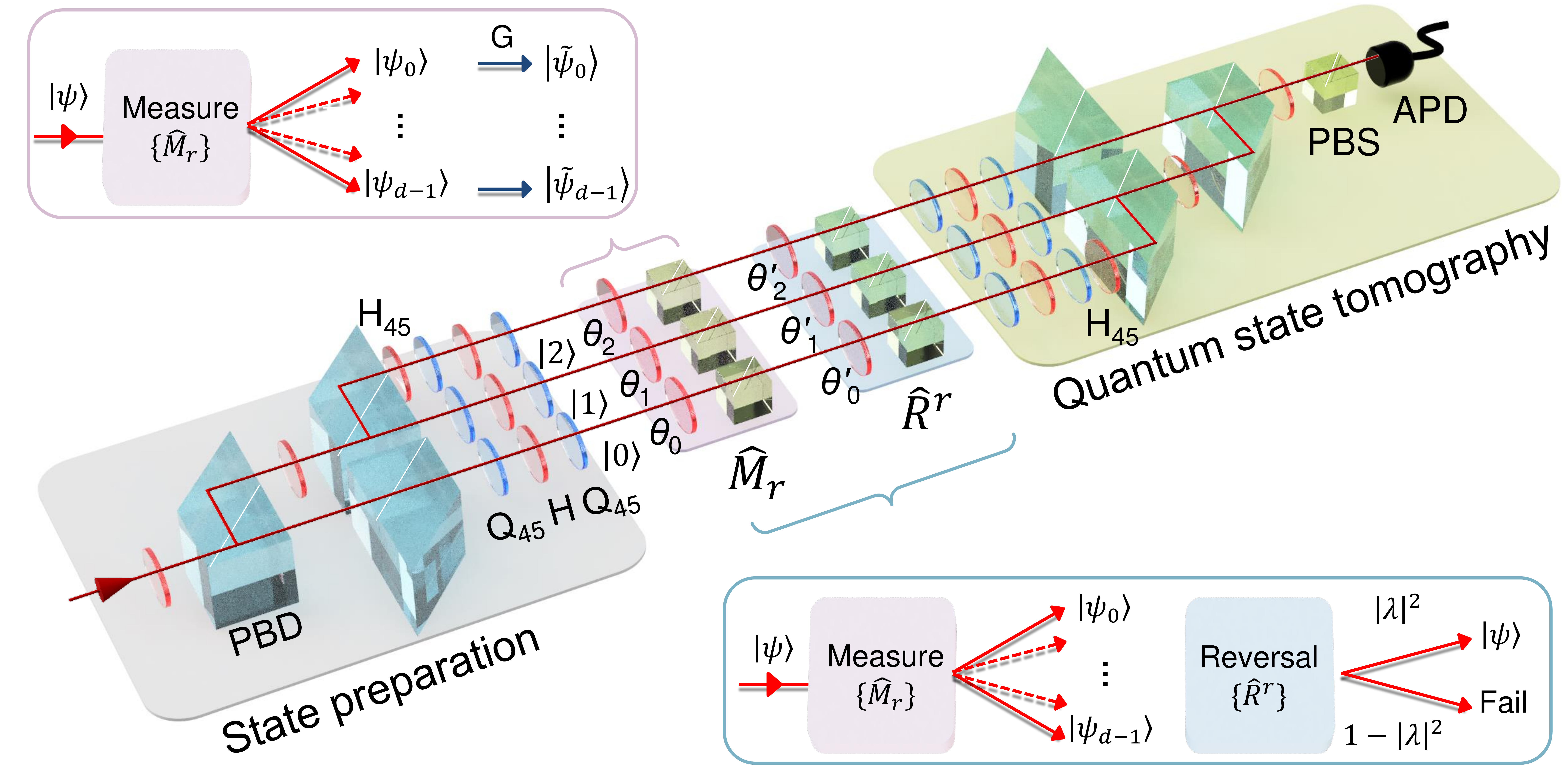} 
\caption{{\bf Schematic of the experimental set-up.} A heralded single-photon qutrit state is prepared to be $|\psi\rangle = a_0|0\rangle+e^{i\phi_{1}}a_1|1\rangle+e^{i\phi_{2}}a_2|2\rangle$ by using HWPs, QWPs, and PBDs. Quantum measurement and reversing operators are realized using a set of three HWPs and three PBSs. By controlling the angles $\theta_i$ of HWPs, a generalized qutrit measurement operator  can be implemented as $\hat{M}_r=\cos 2 \theta_{0} |0\rangle\langle0|+\cos 2 \theta_{1} |1\rangle\langle1|+\cos 2 \theta_{2} |2\rangle\langle2|$ and the corresponding reversal operator can be obtained by the same way as $\hat{R}^{r}=\cos2 \theta_{0}^{'} |0\rangle\langle0|+\cos2 \theta_{1}^{'} |1\rangle\langle1|+\cos 2\theta_{2}^{'} |2\rangle\langle2|$. In our scheme a signal photon passes through the same number optical components, when it propagates through any path mode, so that the difference of optical path length among all the modes is negligible. Note that the relative phases among each path ($\phi_1$ and $\phi_2$) are stabilized without any active stabilization method. See Supplementary Note for the detailed experimental information. (PBD: Polarizing beam displacer, H$_{45}$: Half-wave plate fixed at $45^\circ$, H: Half-wave plate, Q: Quarter-wave plate, PBS: Polarzing beam splitter, APD: Avalanche photo diode)} 
\label{fig:fig2}
\end{figure*}

A quantum measurement can thus split the information of a quantum state into three parts, i.e., the extracted $G$, transmitted (undisturbed) $F$, and reversible $R$ information (see Fig.~\ref{fig:fig1}{\bf b}). These quantities characterize a quantum measurement. 
For example, a von Neumann measurement enables to extract the maximum information $G=2/(d+1)$, while disturbing the state maximally $F=2/(d+1)$ and is thus irreversible $R=0$. On the other hand, a unitary operation extracts no information $G=1/d$ without disturbing the measured system $F=1$ and is deterministically reversible $R=1$. 

The amounts of $G$, $F$, $R$ are fundamentally limited as the total information of an unknown quantum state never increase, which can be understood as a manifestation of the no-cloning theorem \cite{Wootters}. Each of their quantities is thus expected to be related in a trade-off manner.
The trade-off relation between information gain and disturbance ($G$-$F$) was first derived in Ref.~\cite{Banaszek01}. Later, information gain and reversibility were shown to be also in the trade-off relation ($G$-$R$) \cite{Cheong12}. Recently, a global trade-off relation of the three were derived ($G$-$F$-$R$) which has the upper bound
\begin{eqnarray}
\label{eq:t1}
\sqrt{{F}-\frac{1}{d+1}}&\leq& \sqrt{{G}-\frac{1}{d+1}} + \sqrt{\frac{R}{d(d+1)}} \\
\nonumber
&&\hspace{3mm}+\sqrt{(d-2)\bigg(\frac{2}{d+1}-{G}-\frac{R}{d(d+1)}\bigg)},
\end{eqnarray}
as well as the relation between ($F$-$R$) \cite{SWLEE20} (see Method for the detailed forms of other trade-off relations). 

In what follows, we shall explore information contents by applying quantum measurements given in the form of
\begin{equation}
\hat{M}_r \equiv \hat{V}_r \big(\lambda^{r}_0 \ket{0}\bra{0} +  \lambda^{r}_1 \ket{1}\bra{1} + \lambda^{r}_2 \ket{2}\bra{2} \big) \hat{U}_r,
\label{eq:3}
\end{equation}
satisfying $\sum_r \hat{M}_r^{\dag}\hat{M}_r = I$ (we set $\hat{V}_r=\hat{U}_r=I$ for simplicity), where different inputs of $\lambda^r_{i=0,1,2}$ determines the type of quantum measurement. By exploring $G$, $F$ and $R$ by changing the type and strength of quantum measurements, we aim to demonstrate complete trade-off relations and verify the tightness and optimality of the global trade-off relation that emerges in multi-dimensional Hilbert space.


\begin{figure*}[t]
\includegraphics[scale=0.3]{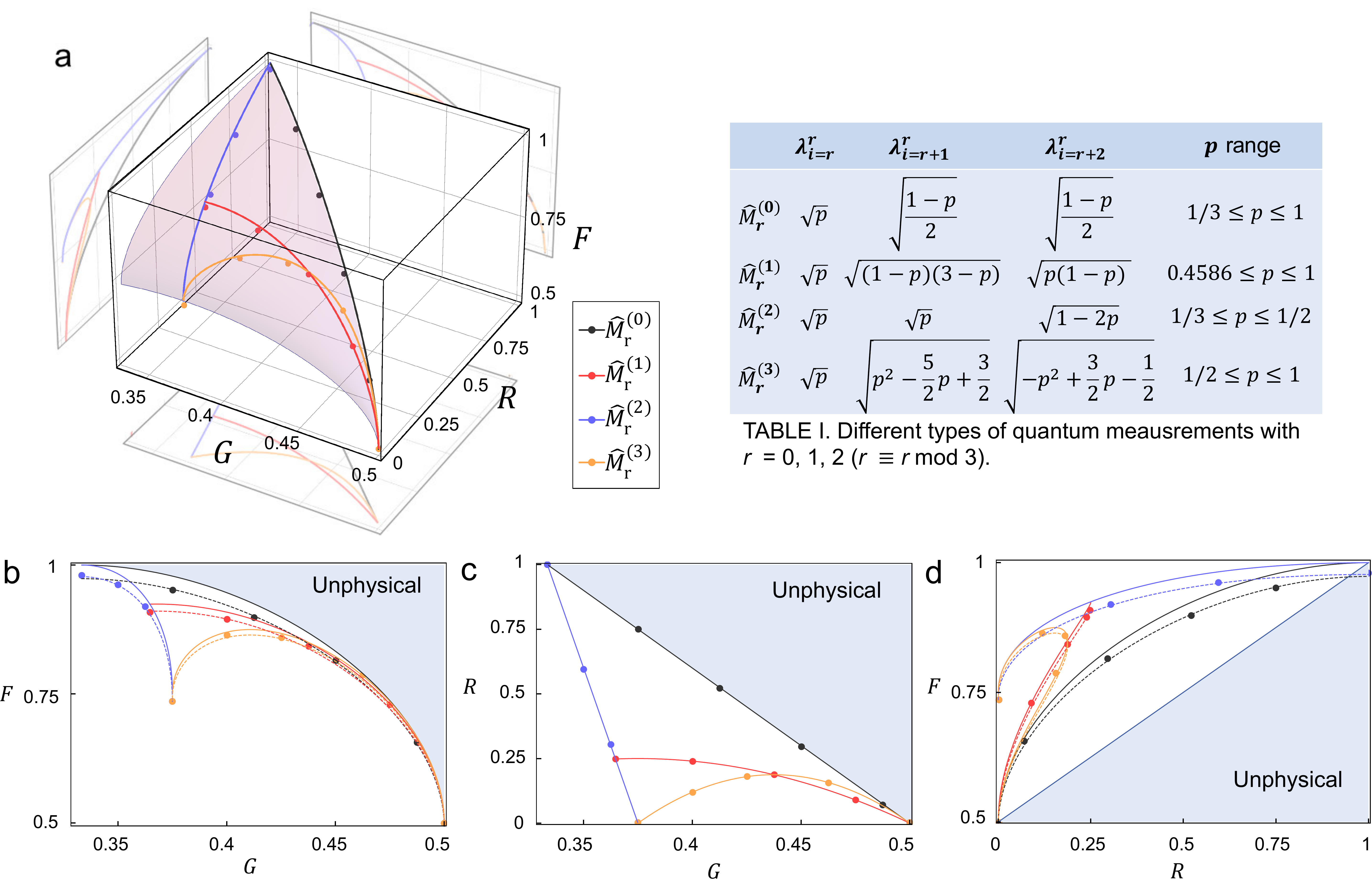} 
\caption{{\bf Complete trade-off relations in quantum measurement.} {\bf a,} Experimentally obtained $G$, $F$, $R$ are plotted for different quantum measurements listed in Table I. Black, red, blue, and yellow dots refer to experimental results for quantum measurement $\hat{M}_{r}^{(t)}$ with $t=0,1,2,3$ and different $p$, respectively, and solid lines represent the expectation value evaluated by continuously changing $p$ for an ideal $\hat{M}_{r}^{(t)}$. The upper bound of the global trade-off relation $G$-$F$-$R$ is represented as a surface. Here, the results of all the measurements $\hat{M}_{r}^{(t)}$ are on the surface. Each pairwise relations are obtained by projecting onto {\bf b}, $G$-$F$, {\bf c,} $G$-$R$, and {\bf d,} $F$-$R$ planes, respectively. The shaded region indicates unphysical regions due to the bound of each pairwise relations. $\hat{M}_{r}^{(0)}$ saturates the upper bounds of both $G$-$F$ and $G$-$R$ for any $p$, while the other measurements cannot saturate none of the pairwise relations except when representing a von Neumann projection or a unitary operator. Note that the bound of $F$-$R$ is only saturated by a von Neumann projection or a unitary operator \cite{SWLEE20}. The dashed lines represent the fitted trade-off relations assuming a non-ideal input state $\hat{\rho} (e) = e|\psi \rangle \langle \psi |+(1-e)I/3$. The error parameters $e$ are obtained to be $e=0.960$, $e=0.977$, $e=0.966$ and $e=0.980$ for $\hat{M}_{r}^{(0)}$, $\hat{M}_{r}^{(1)}$, $\hat{M}_{r}^{(2)}$, and $\hat{M}_{r}^{(3)}$, respectively. The error bars represent one standard deviation obtained by performing 100 Monte-Carlo simulation runs by taking into account of the Poissonian photon counting statistics. The error bars are too small to be visible.}
\label{fig:fig3}
\end{figure*}

To implement quantum measurements in Eq.(\ref{eq:3}), we employ a photonic path-encoded qutrit system. As illustrated in Fig.~\ref{fig:fig2}, we prepare an arbitrary qutrit state $|\psi\rangle = a_0|0\rangle+e^{i\phi_{1}}a_1|1\rangle+e^{i\phi_{2}}a_2|2\rangle$, where $\sum_ia_{i}^{2}=1$ (See Method for details). The prepared states undergo quantum measurement $\hat{M_{r}}$ or both quantum measurement $\hat{M_{r}}$ and reversing operation $\hat{R}^r$. To evaluate $G$ and $F$ for a qutrit system, we perform the symmetric and informationally complete positive-operator-valued measures (SIC-POVM) with nine states \cite{HTLim12}. POVMs for $\hat{M_{r}}$ and $\hat{R}^{r}$ are realized with a set of half-wave plates (HWPs) and polarizing beam splitters (PBS). Since PBS transmits the horizontally polarized photons and reflects the vertically polarized photons, the transmission ratio can be varied by adjusting the angle of HWP. The transmission amplitude for each path is modified by using a set of HWP and PBS. Different types of quantum measurement operator in Eq.~(\ref{eq:3}) can be realized by adjusting the angle $\theta_i$ of HWP as $\lambda^{r}_i=\cos2\theta_{i}$ with $i=0, 1, 2$ as shown in Fig.~\ref{fig:fig2}.
The reversing operator $\hat{R}^r$ can be implemented by the same way as $\hat{M_{r}}$ using a set of HWPs of $\theta'_{i}$ and PBSs.

We realize different quantum measurements $\hat{M}^{(t)}_{r}$ with $t=0,1,2,3$ (the explicit forms are given in the Table in Fig.~\ref{fig:fig3}) by varying the measurement strength parameterized by $p$. We evaluate $G$ and $F$ for $\hat{M}^{(t)}_{r}$ with specific $p$ by analyzing the quantum state tomography (QST) for nine pure states of SIC-POVM. $R$ is evaluated by analyzing the final state after both $\hat{M}_r$ and $\hat{R}^{r}$ are performed. See Supplementary Note for the detailed information on how to extract $G$, $F$, and $R$ quantities from the experimental data. To verify that the initial state is retrieved after reversing operation, we perform quantum process tomography (QPT) for analyzing the realized operation. Detailed information on QPT results are provided in Supplementary Note. 


Figure~\ref{fig:fig3} presents the experimental results. Three information contents are investigated for quantum measurements $\hat{M}^{(t)}_r$ with $t=0,1,2,3$ and different $p$. The obtained result $(G, R, F)$ for each $\hat{M}^{(t)}_r$ and $p$ is plotted as a dot in Fig.~\ref{fig:fig3}{\bf a}, where the solid line represents the theoretical expectation value evaluated by continuously changing $p$. We can observe that the amount of $G$, $F$, and $R$ tend to vary in a trade-off manner and global exchanges occur among them as $p$ changes. The amount of disturbance $1-F$ and information gain $G$ by $\hat{M}^{(0)}_r$, $\hat{M}^{(1)}_r$, $\hat{M}^{(2)}_r$ exhibit monotonic increases as increasing $p$, while the reversibility $R$ decreases. $\hat{M}^{(3)}_r$ draws a non-trivial quadratic curve of information exchanges as $\lambda^r_{i=1,2}$ is given as a quadratic function of $p$. The amount of $G$, $F$, and $R$ depending on $p$ are plotted in Supplementary Note for each quantum measurement $\hat{M}^{(t)}_r$. While these demonstrate triplewise quantitative links among $G$, $F$, and $R$ for different types of quantum measurements, all the obtained results hold a certain upper limit represented as a shaded surface in Fig.~\ref{fig:fig3}{\bf a}, which exactly meets the bound of Eq.(\ref{eq:t1}). Notably, $G$, $F$, and $R$ for $\hat{M}^{(t)}_r$ with $t=0,1,2,3$ are always on the surface irrespective of $p$, i.e., saturate the global trade-off relation Eq.(\ref{eq:t1}). 

To investigate the pairwise trade-off relations $G$-$F$, $G$-$R$ and $F$-$R$, we project the 3D plot in Fig.~\ref{fig:fig3}{\bf a} onto each corresponding plane. The obtained results are plotted in Fig.~\ref{fig:fig3}{\bf b}, {\bf c}, and {\bf d}, where the border to the shaded region indicates the theoretical bounds of $G$-$F$, $G$-$R$, and $F$-$R$ relation, respectively (see Table~\ref{tab:table2} in Method). We can observe the quantitative links between two selected pair of information contents as varying $p$. At two extremal points in each plot, i.e., when representing a von Neumann projection (e.g.,~$\hat{M}^{(0)}_r$ with $p=1$) or a unitary operation (e.g.,~$\hat{M}^{(2)}_r$ with $p=1/3$), all measurements reach the upper bounds of the pairwise trade-off relations. Otherwise, each measurement $\hat{M}^{(t)}_r$ show different tendencies. For example, $\hat{M}^{(0)}_r$ enables to reach the upper bound of $G$-$F$ and $G$-$R$ trade-off relations for any value of $p$. 
On the other hand, none of $\hat{M}^{(1)}_r$, $\hat{M}^{(2)}_r$, $\hat{M}^{(3)}_r$ allows us to attain $G$, $F$, $R$ saturating the bounds of the pairwise relations.

Remarkably, our result shows that a certain class of quantum measurements that saturates neither of the pairwise trade-off relations can saturate the global trade-off relation. This in turn indicates that the global trade-off relation is tighter than any of the pairwise trade-off relations.
We note that optimizing quantum measurement can be generally aimed at extracting information without any loss of information. Previously, an optimal quantum measurement has been required to saturate either information-disturbance $G$-$F$ \cite{Sciarrino06} or information-reversibility $G$-$R$ trade-off relation \cite{HTLim14}. Within this condition, $\hat{M}^{(0)}_r$ is an optimal quantum measurement that can extract information $G$ maximally for a fixed amount of $F$, but $\hat{M}^{(1)}_r$ is non-optimal as shown in Fig.~\ref{fig:fig3}{\bf b}. However, our result is a clear evidence that, when $G$, $F$, $R$ are simultaneously taken into account, there exist optimal quantum measurements beyond the previously identified ones. In fact, we can reverse $\hat{M}^{(1)}_r$ to faithfully recover the original quantum state with probability $R$, which accounts for the gap between $F$ and $G$ regarded as a missing part of information previously.
Therefore, we can generally define an {\it optimal quantum measurement} as a measurement that transfers the total information of a quantum state into $G$, $F$, and $R$ without unaccounted part of information as illustrated in Fig.~\ref{fig:fig1}{\bf b}, i.e.,~{\it a quantum measurement inherently preserving quantum information}. An optimal quantum measurement thus saturates at least one of the trade-off relations. See Method for the saturation conditions of trade-off relations. 
As a result, all the measurements $\hat{M}^{(t)}_r$ with $t=0,1,2,3$ realized in our experiment are optimal by saturating the global tight trade-off relation $G$-$F$-$R$.

\begin{figure}[t]
\includegraphics[scale=0.38]{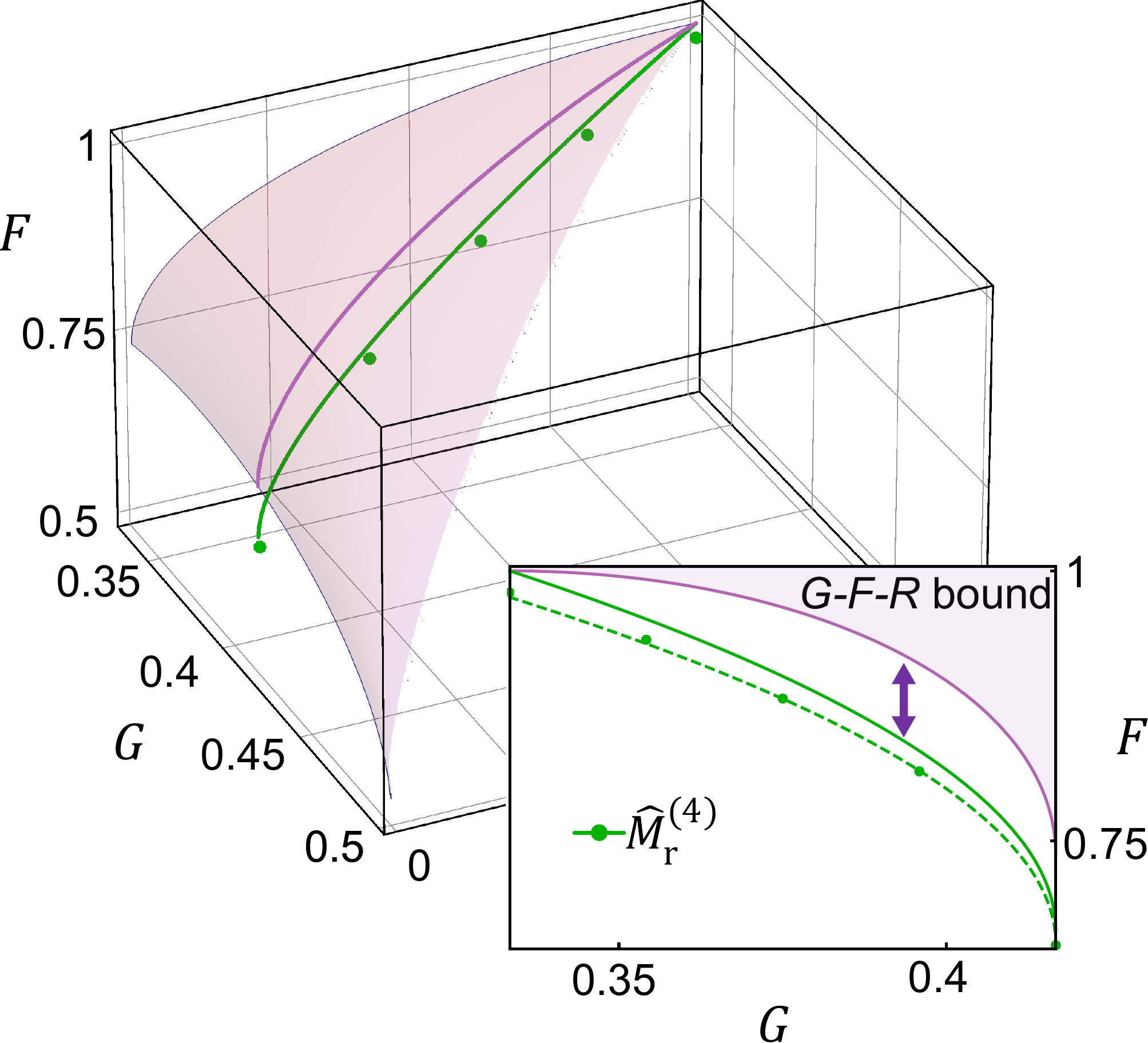}
\caption{{\bf Information loss by a weak quantum measurement.} Experimentally obtained $G$, $F$, and $R$ for a quantum measurement $\hat{M}^{(4)}_r$ are plotted by dots. The green solid line represents the changes of $G$, $F$, and $R$ obtained assuming ideally performed $\hat{M}^{(4)}_r$ in the region $0 \leq p \leq 1$, while the purple solid line corresponds to the bound of the global trade-off relation.
$\hat{M}^{(4)}_r$ cannot reach the upper bound of the global trade-off relation except when it represents a unitary operator $p=0$. The green dashed line refers to results obtained assuming a non-ideal input state with $e=0.963$. Experimental results for pairwise trade-off relations for $\hat{M}^{(4)}_r$ are provided in Supplementary Note.} 
\label{fig:fig4}
\end{figure}

A loss of information may arise in quantum measurement due to the effect of noise, ignorance in estimating the quantum state, or inherent non-optimality: (i) In our experiment, we consider the effect of noise to the input states as $\hat{\rho}(e)=e\ket{\psi}\bra{\psi}+(1-e)I/3$, resulting in dashed lines in Fig.~\ref{fig:fig3}. We can also take into account noise that makes the final output state in Fig.~\ref{fig:fig1}{\bf b} deviated from the original input, i.e., $\hat{R}^{r}\hat{M}_r\ket{\psi} \propto \ket{\psi'} \neq \ket{\psi}$, which reduces $R$ \cite{SWLEE20}. So, the effect of noise generally brings about a degradation either the amount of $G$, $F$, $R$ so that none of the trade-off relations can be saturated. 
(ii) Ignorance and inefficiency when estimating a quantum state from measurement data may directly reduce the amount of information gain $G$ \cite{Chen14,HTLim14}. (iii) Interestingly, it turns out that the form of quantum measurement itself can also induce information loss. For example, consider a weak quantum measurement $\hat{M}^{(4)}_r$ defined by $\hat{M}_{0}^{(4)}=|0\rangle\langle 0|+\sqrt{1-p} |1\rangle\langle 1|+|2\rangle\langle 2|$ and $\hat{M}_{1}^{(4)}=\sqrt{p} |1\rangle\langle 1|$ for $0 \leq p \leq 1$. We experimentally obtain $G$, $F$, and $R$ by changing $p$ and find that none of the trade-off relations can be saturated (see Fig.~\ref{fig:fig4}) (except when $p=0$). $\hat{M}^{(4)}_r$ may be inherently non-optimal as there exists a part of information that is not changed into any of $G$, $F$, $R$, even if the measurement is performed perfectly without noise and ignorance (see Method).

From a fundamental point of view, our result is a first experimental proof of the global trade-off relation among the extracted, disturbed, and reversible information. We also realized quantum measurement preserving quantum information in the sense that all information of a quantum state transferred to $G$, $F$, and $R$ without any missing part. This provides a general condition of an optimal quantum measurement and verifies the information conservation law in quantum measurement, which constitutes a basis of secure quantum information processing. We observed that the global trade-off relation is tighter than any pairwise trade-off relations. It implicates that the conventional wisdom, 
on the information-disturbance trade-off should be rephrased more rigorously into `the more information is obtained by quantum measurement, the more the state is disturbed or less recoverable.' In addition, our work raises a fundamental question on the information loss by an inherently non-optimal quantum measurement, which may provide a deep insight on the quantum to classical transition and loss of quantumness in quantum measurement \cite{Johannes07,Jeong14}. Further studies on information loss as increasing the dimensionality of quantum measurement may be valuable. 

While our demonstration is executed based on photonic qutrits ($d=3$), the results are valid for arbitrary dimensional systems. We have observed that the role of the reversibility $R$ emerges in multi-dimensional Hilbert space, while the global trade-off relation reduces to the information-disturbance relation $G$-$F$ when $d=2$ \cite{SWLEE20}. Our results may be thus particularly useful in high-dimensional quantum information processing \cite{Erhard20}. We have demonstrated different types of optimal quantum measurements. As analyzed in Ref.~\cite{SWLEE20}, such measurements can be classified into different sets according to which trade-off relation they saturate. Thus, different optimal quantum measurements may suit to different applications. In order to estimate or discriminate quantum states, an optimal measurement that leads to maximum information gain $G$ with minimal disturbance (i.e.,~maximum $F$) may be required \cite{HTLim14}. To transfer or stablize qubits e.g.,~in quantum teleportation \cite{AdvancedTele} or quantum error correction \cite{QEC,Koashi99}, maximum reversibility $R$ and minimal information gain $G$ may be desirable as a reversing operation plays an important role to recover the input information. Implementing optimal quantum measurements in specific quantum information protocols may be a next step of research.

We hope that our work deepen our understanding on quantum measurement and would find wider applications in current developments of quantum technologies. 


\section*{Methods}
{\bf Photonic path mode qutrit system.}
We used a CW single frequency laser operating at a center wavelength of 405.5 nm. A pair of photons with a center wavelength of 811 nm is generated via spontaneous parametric down conversion (SPDC) process by pumping a 20 mm long type-II periodically poled KTiOPO$_{4}$ (PPKTP) crystal with the poling period of 10 $\mu$m. The idler photon was detected at a trigger detector to prepare the signal photon as a heralded single-photon state, and the signal photon is delivered to the experimental setup shown in Fig.~\ref{fig:fig2}. Since a polarizing beam displacer (PBD) transmits (reflects) horizontal (vertical) polarization states, by adjusting the angles of HWPs in front of the PBD, a signal photon can have three possible path modes, lower $|0\rangle$, middle $|1\rangle$, and upper $|2\rangle$ path modes. Polarization of photons in all path modes are arranged to be a horizontal polarization state ($\left | H \right \rangle$) by locating a HWP with 45$^{\circ}$ (H$_{45}$) at the upper path $|2\rangle$, see Fig.~\ref{fig:fig2}. The relative phases among three path modes can be controlled by a set of two quarter-wave plates (QWPs) fixed at 45$^{\circ}$ and HWP with an angle of $\alpha$ without changing the polarization state of the signal photons. By adjusting the HWP angle $\alpha$ between two QWPs, one can implement relative phase $\phi=2\alpha$. Then, one can prepare an arbitrary qutrit state $|\psi\rangle = a_0|0\rangle+e^{i\phi_{1}}a_1|1\rangle+e^{i\phi_{2}}a_2|2\rangle$, where $\sum_ia_{i}^{2}=1$. 

{\bf Bounds of trade-off relations.}
The bounds of pairwise trade-off relations between two information contents are listed in Table~\ref{tab:table2}. Quantum measurements that saturate different trade-off relations have different forms of the measurement operators. The singular values $\lambda^{r}_i$ determines the form of quantum measurement. We can define a vector of singular values for each $i$ as $\vec{v}_i=(\lambda^{r=0}_i, \ldots , \lambda^{r=N}_i)$. 
A quantum measurement saturates the upper bound of the global trade-off relation Eq.(\ref{eq:t1}) if and only if all $\vec{v}_i$ for $i=0,\cdots,d-1$ are collinear and satisfy $|\vec{v}_1|=\cdots=|\vec{v}_{d-2}|$ \cite{SWLEE20}. Similarly, a quantum measurement saturates the upper bound of $G$-$F$ if and only if all $\vec{v}_i$ for $i=0,\cdots,d-1$ are collinear and satisfy $|\vec{v}_1|=\cdots=|\vec{v}_{d-1}|$ \cite{Banaszek01}. On the other hand, the bound of $G$-$R$ is saturated when the measurement operators satisfy $\hat{M}^{\dag}_r\hat{M}_r=a_r\ket{i_r}\bra{i_r}+b_r\hat{\openone}$ with non-negative $a_r$ and $b_r$ \cite{Cheong12}. The bound of $F$-$R$ is saturated if and only if the measurement is a von Neumann projection or unitary operator \cite{SWLEE20}.

\begin{table}[h]
\setcounter{table}{1}
\caption{\label{tab:table2} The pairwise trade-off relations  \cite{Banaszek01,Cheong12,SWLEE20}.}
\begin{ruledtabular}
\begin{tabular}{l|c}
&Trade-off relation \\
\hline
$G$-$F$ &  $\sqrt{F-\frac{1}{d+1}}\leq \sqrt{G-\frac{1}{d+1}} +\sqrt{(d-1)(\frac{2}{d+1}-{G})}$  \\
$G$-$R$ & $d(d+1){G}+(d-1){R} \leq 2d$ \\
$F$-$R$ & $2\leq (d+1){F}-(d-1){R}$ \\
\end{tabular}
\end{ruledtabular}
\end{table}

{\bf A quantum measurement inherently losing information.}
We consider a quantum measurement $\hat{M}_{r}^{(4)}$ described by operators $\hat{M}_{0}^{(4)}=|0\rangle\langle 0|+\sqrt{1-p} |1\rangle\langle 1|+|2\rangle\langle 2|$ and $\hat{M}_{1}^{(4)}=\sqrt{p} |1\rangle\langle 1|$ for $0 \leq p \leq 1$, which satisfy the completeness condition $\hat{M}_{0}^{(4)} \hat{M}_{0}^{(4)\dag}+\hat{M}_{1}^{(4)} \hat{M}_{1}^{(4)\dag} =\hat{I}$. We explore the amount of information gain, disturbance and reversibility of $\hat{M}_{r}^{(4)}$ by changing $p$. We find that the obtained $G$, $F$, and $R$ do not reach the upper bound of the global trade-off relation $G$-$F$-$R$ (except when $p=0$) as shown in Fig.~\ref{fig:fig4}. These also not saturate any of the pairwise trade-off relations (see Supplementary Note).
We can also theoretically evaluate the maximum amount of information gain, disturbance and reversibility of $\hat{M}_{r}^{(4)}$, resulting in $G=(4+p)/12$, $F=(2+\sqrt{1-p})/3$, and $R=1-p$ given as a function of $p$. By these, we can easily find that any of the trade-off relations is not saturated except when it represents a unitary operation ($p=0$). This thus holds for an ideal case when the measurement is performed without any noise and ignorance. As a result, there exists a part of losing information due to the inherent non-optimality of $\hat{M}_{r}^{(4)}$.




\section*{Acknowledgements}
This work was supported by the National Research Foundation of Korea (NRF) (2019M3E4A1079777, 2019M3E4A1078660, 2020M3E4A1079939, 2021R1C1C1003625), the Institute for Information and Communications Technology Promotion (IITP) (2020-0-00947, 2020-0-00972), and the KIST research program (2E31021).


  


\end{document}